# COMMON AND DIFFERENT FEATURES BETWEEN THE BEHAVIOR OF CHAOTIC DYNAMICAL SYSTEMS AND THE $1/f^\alpha$ -TYPE NOISE

M. K. Koleva
*Institute of Catalysis, Bulgarian Academy of Science, 1113 Sofia, Bulgaria*
E-mail: mkoleva@bas.bg

V. Covachev
*Institute of Mathematics and Informatics, Bulgarian Academy of Science, 1113 Sofia, Bulgaria*
E-mail: matph@math.bas.bg

**Abstract**

The major goal of the present paper is to find out the manifestation of the boundedness of fluctuations. Two different subjects are considered: (I) an ergodic Markovian process associated with a new type of large-scaled fluctuations at spatially homogeneous reaction systems; (ii) simulated dynamical systems that possess strange attractors. Their common property is that the fluctuations are bounded. It is found out that the mathematical description of the stochasticity at both types of systems is identical. Then, it is to be expected that it exhibits in certain common features whose onset is the stochasticity, namely: (I) The power spectrum of a time series of length $T$ comprises a strictly decreasing band that uniformly fits the shape $1/f^{\alpha(f)}$ where $\alpha\left(\frac{1}{T}\right) = 1$ and $\alpha(f)$ strictly increases to the value $\alpha(\infty) = p$ $(p > 2)$ as $f$ approaches infinity. Practically, at low frequencies this shape is $1/f$ -like with high accuracy because the deviations of the non-constant exponent $\alpha(f)$ from 1 are very small and become even smaller as the frequency tends to $1/T$. The greatest advantage of the shape $1/f^{\alpha(f)}$ is that it ensures a finite variance of fluctuations. (ii) It is found out that the structure of a physical and a strange attractor is identical and they are non-homogeneous. The explicit expression of the phase space density law is derived. (iii) The Kolmogorov entropy is positive and finite. Thus, the admitted so far as the distinctive for the deterministic chaos properties are shared by $1/f^{\alpha(f)}$ noise. It is found out that the only practical discrimination criterion is the superimposing of a discrete and a continuous band of the shape $1/f^{\alpha(f)}$ at certain values of the control parameters.

## 1. Introduction

The identification of chaotic dynamical systems that possess strange attractors is an important research topic from the viewpoint of chaotic time series analysis. It is well known fact that they have three distinctive properties: (I) the power spectra are continuous strictly decreasing bands; (ii) the correlation dimension of the strange attractors is non-integer; (iii) the Kolmogorov entropy is positive and finite. However, recent results [1] on simulation of chaotic dynamical systems yield $1/f^\alpha$-type tails in the power spectra. The significance of this result is revealed in the fact that the chaotic dynamical systems are bounded while the variance of the fluctuations, calculated on the ground of the spectral density, is infinite. The problem has also a physical insight: there is a large variety of open spatially homogeneous reaction systems which exhibit $1/f^\alpha$-type tails in their power spectra. The infinite variance makes most probable large fluctuations enough to carry the system beyond the physical thresholds of stability. Therefore, any system would blow up or get extinct in a finite time interval. However, it does not happen. Moreover, the $1/f^\alpha$-type behavior is spanned over several dozens of order. The problem becomes even more complicated because, as a rule, these systems are governed by special precautions such that fluctuations are small. Then, their macroscopic evolution is considered by systems of ordinary differential equations. So, what is the physical source of the $1/f^\alpha$-type behavior of the spectral density? What is the relation to the chaotic dynamical systems? Can reconciliation between the boundedness of a system and the $1/f^\alpha$-type behavior of the spectral density be achieved? In the course of revealing these problems the major question arises whether the shape of the spectral density is exactly $1/f^\alpha$ or it is a good low frequency approximation to something else. In the recent experiment on catalytic oxidation of HCOOH (formic acid) over supported Pd (palladium) catalyst [2] the shape of the spectral density of the catalyst temperature variations in the time course has been studied. 80 time series are recorded at different values of the control parameters over 3 modifications of the catalyst. The sampling rate is 2 points per second. It has been found out that the shape of the power spectra obeys the law $1/f^{\alpha(f)}$ where $\alpha(f)$ is non-constant and strictly increases from $\alpha(f) \approx 1.1$ at low frequency limit to $\alpha(f) \approx 1.2$ at higher frequencies. These values are typical for all the power spectra. The smooth increasing deviations of $\alpha(f)$ from 1 has been established by scanning the coordinates of the consecutive components of each power spectrum. They become even visible for higher frequencies (Figs. 1,2).

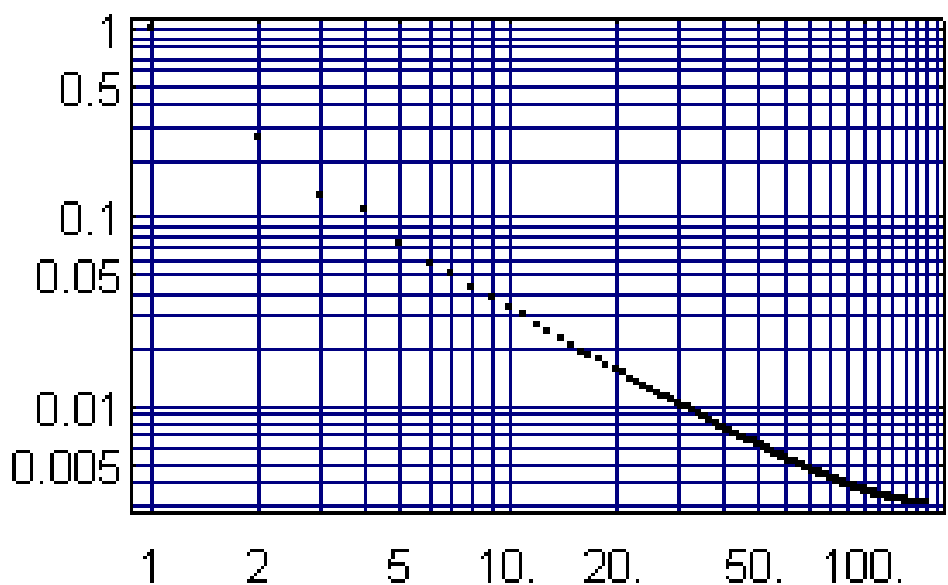

Fig. 1. A power spectrum in relative units of the catalyst temperature variations (K) in the course of time (sec) at the oxidation of HCOOH at a single stable state

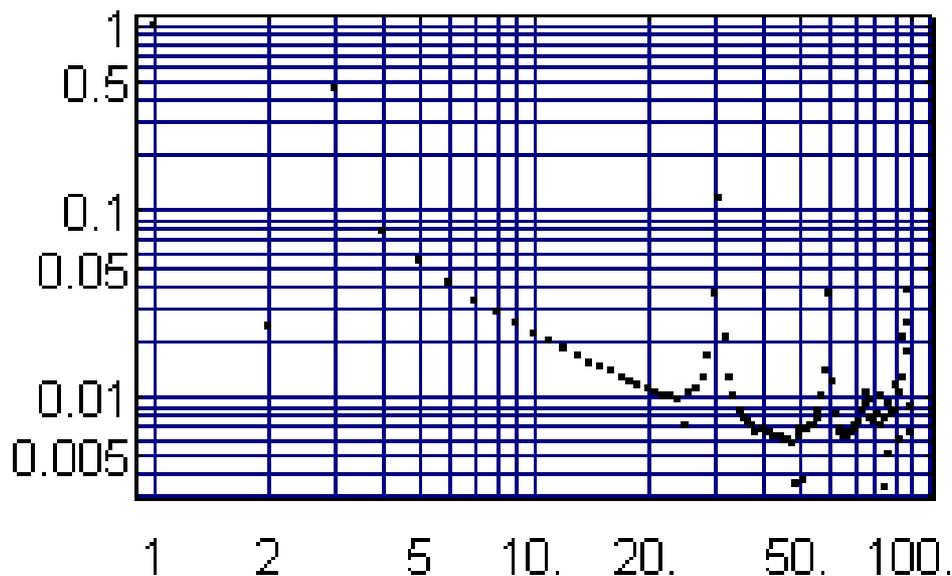

Fig. 2. A power spectrum in relative units of the catalyst temperature variations (K) in the course of time (sec) at the oxidation of HCOOH at a limit cycle

Another curious result is that at certain values of the control parameters a discrete band is superimposed on the continuous one (Fig. 2). This is also a rather controversial result because there is no such class of differential equations the power spectra of whose solutions can comprise both a discrete and a continuous band. The problem becomes even more complicated since both bands are of the same order of magnitude. Hence the assumption that the discrete band originates from a deterministic process, *e.g.*, a limit cycle, gives rise to the question what the origin of the continuous one is? Obviously, it is not a chaotic dynamical system since if this were the case: (I) the continuous band could not coexist with a discrete one; (ii) the continuous band would appear only at certain values of the control parameters.

It is well known also that the observed properties of the power spectra are not typical for the established so far sources of large-scaled fluctuations. Moreover, their influence is enhanced to macroscopic effects only near the points of macroscopic instability, *e.g.*, bifurcation points.

The robustness of the observed continuous band to dynamical regimes and the same order of the discrete and the continuous band suggest that the source is a new type of large-scale fluctuations (noise). Recently one of us proposed a new physical mechanism [3] that brings about a new type noise called by the author diffusion-induced noise (according to its origin). This mechanism is typical for all the surface reactions and is insensitive to the details of the reaction mechanism. Its most important for the further consideration property is that it constitutes a continuous ergodic Markovian process where the fluctuations are bounded. It is worth noting that this process is different from the Ornstein-Uhlenbeck one. The details of the physical mechanism are presented in &2.

It is shown in &3 that the simulation produces a stochastic behavior of the r.h.s. of the chaotic dynamical systems. The small fluctuations that originate from the inevitable round-off increase in the course of time to values compatible to that of the r.h.s. due to at least one positive Lyapunov coefficient. The amplitude of the fluctuations is bounded since each phase trajectory is confined in a strange attractor. It is shown that the mathematical description of this stochasticity is identical to that of the diffusion-induced noise.

So, we have two types of bounded systems with the similar low frequency limit of the power spectra, namely $1/f^\alpha$-type decay: simulated dynamical systems that possess strange attractors and a continuous ergodic Markovian process with bounded fluctuations. Before establishing their common properties we present our physical arguments about the boundedness of fluctuations. The basis of our further considerations is the assertion that a stable long-term evolution of any system is possible if and only if involving locally a finite energy and/or matter at any instant drives the fluctuations. The limit values are determined so that the system stays permanently within the thresholds of stability of the system. It is obvious that the limit values are specific to the system but they are *always finite*. These statements constitute the physical meaning of the notion boundedness of fluctuations. The driving mechanism of the diffusion-induced noise corresponds to this assumption since it involves locally finite energy and matter to produce a fluctuation so that the system stays permanently stable. It is clarified in &2. However, the idea of the boundedness of fluctuations is more general since it can arise as a result of variety of reasons. For example, the confinement of each phase trajectory in a strange attractor at chaotic dynamical systems is due to the structure of the r.h.s. of some systems of differential equations. The driving mechanism of fluctuations at simulation, presented in &3, is of nature different from that of the diffusion-induced noise.

The major goal of the present paper is to elucidate whether there are common properties of the simulated chaotic dynamical systems and the diffusion-induced noise and how they are related to the boundedness of the systems. We evaluate analytically the shape of the spectral density, the correlation dimension of the strange and physical attractors and the Kolmogorov entropy. The invariant set designed by the admissible values of the fluctuations is called next physical attractor. Surprisingly, it is found out that the structure of the physical and strange attractor is identical, the attractors are non-homogeneous and the correlation dimension is non-integer. It is found out also that the Kolmogorov entropy is positive and finite for both types of systems. These results are presented in &5. The shape of the power spectra is established in &4. It is found out that the power spectrum of any bounded irregular time series of finite length $T$ comprises a strictly decreasing band which uniformly fits the shape $1/f^{\alpha(f)}$ where $\alpha(f) \in [1, p]$, where $p$ is arbitrary but $p > 2$. $\alpha(f)$ strictly increases starting from the value $\alpha = 1$ at $f = 1/T$ up to $\alpha = p$ as $f$ approaches infinity. In turn, this shape provides finite variance of the fluctuations, calculated on the ground of the spectral density. Thus, stable long-term evolution of bounded systems is ensured. It is worth noting that these properties are due only to the boundedness of fluctuations and to the Markovian property of the fluctuation appearance in the course of time. The major difference between this shape and the $1/f^\alpha$ ($\alpha$ constant) is that the latter yields infinite variance of fluctuations no matter what the value of $\alpha$ is. Therefore it cannot ensure a stable long-term evolution of any physical system.

The difference appears in the coexistence of a discrete and a continuous band. It turns out that a power spectrum of the diffusion-induced noise can comprise both a discrete and a continuous band of $1/f^{\alpha(f)}$ shape whereas only a continuous band of that shape is available for the simulated chaotic dynamical systems. Another difference is that the continuous band of the above shape is persistent at all the values of the control parameters when dealing with diffusion-induced noise while it appears only at certain values of the control parameters at the simulated dynamical systems.

## 2. Physical Mechanism of the Diffusion-Induced Noise

In the Introduction we put forth the hypothesis that the successful explanation of the considered experiment [2] requires a new source of large-scale fluctuations (noise). Before considering the details we briefly outline the established so far sources of noise. The purpose is to elucidate why they fail in the explanation.

*Fluctuating environment* − Coupled to a fluctuating environment, a non-linear open system can display types of behavior that are impossible under corresponding deterministic external constraints. Indeed, when the characteristics of the external fluctuations (i.e., variance, and correlation time) cross certain thresholds, transition phenomena can take place without any change at all in the average state of the environment. This new class of non-equilibrium transitions is called noise-induced transitions [4]. They are characterized by the fact that the system no longer adjusts its macroscopic behavior to the average value of the external constraints.

*Discreteness of the system* − The theory of the many-body systems is built upon the idea that their macroscopic behavior is determined by a limited number of variables, called "relevant" by some authors. Then, on coarse-grained scale, the exact mechanical equations are replaced by stochastic equations. The notion of coarse-graining is considered as averaging of the relevant variables over a finite phase volume. Then, the stochastic behavior of the relevant variables originates in the inevitable interaction with the large set of the remaining degrees of freedom, termed irrelevant. A major property of the fluctuations that arise from this interaction is that they appear as "external" to a given mode. This provides the essential difference of these fluctuations from the "internal" ones, i.e., the fluctuations that arise from the random deviations of the dynamical variables from their average values. The amplitude of the internal fluctuations is of the order of $N^{-1/2}$, where $N$ is the number of the molecules in the system. Thus they are negligible for large systems. The "environment" defines the intensity of the "external" fluctuations and therefore it is independent of $N$. The most famous representative of this kind of fluctuations is the Brownian motion [5,6]. The separation of the reactive collisions of one another results in such kind of noise in chemical systems [7].

*Deviations from the Poissonian distribution* − The most common way of description of the fluctuations in reaction systems is through master equation for the probability distribution. In the frame of this approach, however, the spatial distribution of the reactants does not adjust to the random (Poissonian) one in the presence of any non-linear reaction step. These deviations are responsible for differences between the structure of the phenomenological equations of the evolution and the equations for the first moment of the distribution, (i.e., average values). The influence of the non-Poissonian parts of the higher correlation moments is enhanced in the course of time near a macroscopic instability [8,9].

Summarizing, these sources of noises bring about a macroscopic effect only at certain values of the control parameters (near macroscopic instabilities). Moreover, neither of them gives rise to a continuous band in the power spectra of the output characteristics of any open system. Fluctuations about stable states are mostly small, however, there also occur occasional large fluctuations. Their significance is traced in the fact that they can cause transitions between multiple steady states. This determines the importance of finding out their probability distribution. The approaches to the problem involve solving non-linear master equations for the fluctuation distribution and/or Fokker-Planck equation with different characteristic of noise [10–12]; treating the fluctuations as random perturbations in the analysis of the solutions stability [13]. Recently the eikonal approximation is applied to the problems of large fluctuation [14]. A large class of problems of the fluctuations is deduced by their presentation through stochastic differential equations, in particular of Ito or Stratonovich type (introduction to the problem in [8,15–17]).

Neither of these approaches treats fluctuations as bounded which, however, is a distinctive property of the diffusion-induced noise. Another important property of this noise is that it is persistent even at asymptotically stable states. That is why, it results in new and, to a certain extent, unexpected properties of the macroscopic evolution of a reaction system at steady external constraints.

Next, the physical mechanism of the new type noise is presented for the case when the fluctuations that come from the above sources are negligible, i.e. far from the macroscopic instabilities and unstable macroscopic states.

The theory of the diffusion-induced noise is built upon the idea of a diffusion-induced non-perturbative interaction. The notion of a non-perturbative interaction is defined when a Hamiltonian changes qualitatively under certain perturbation(s). The changes involve opening or closing of output channel(s). These interactions are not exotic occasions but they are even necessary for proceeding of any reaction. To elucidate the statement one should refer to the notion of stoichiometry. When concerning a single reaction event it implies that the latter is initiated if and only if certain species of required types and states form certain local configuration. Then, the Hamiltonian of each of the reactants becomes unstable to the perturbations induced by the other species. The process leads eventually to the formation of the reaction products that are different from the reactants. Therefore, the *ad hoc* assumption implies that the initial Hamiltonian changed qualitatively during reaction (apart from that of a catalyst species). The necessity of a certain local configuration for driving any reaction event is immediately related to the boundedness of matter and energy involved locally. When the Hamiltonian of an interaction changes qualitatively after involving another species and when it happens due to the diffusion of that species, the non-perturbative interaction is called a diffusion-induced one. The latter causes a non-smooth change in the probabilities for each of the output channels. An example of this interaction is presented next.

Next we consider the diffusion-induced non-perturbative interactions at the surface reactions when a surface is exposed to a steady flow of reactants. The purpose is twofold. On the one hand, the oxidation of HCOOH over supported Pd catalyst belongs to this class of reactions. And one of our tasks is the successful explanation of the experiment [2] mentioned in the Introduction. On the other hand, it will be shown next that diffusion-induced non-perturbative interactions appear in the process of chemisorption. The latter is a feature, typical for all surface reactions since a reaction event is available only between already chemisorbed species of required types. Therefore, the noise behavior is insensitive to the details of the reaction mechanism.

The gas/surface system is approximated by the ideal adsorption layer approach introduced by Langmuir. Besides its apparent simplicity it allows the system to be considered as a spatially homogeneous one. The approach is based on the following assumptions:

- A lattice at whose vertices identical active sites are placed presents the catalyst surface. The adsorption and reaction are available only at the active sites;
- the interactions between adjacent chemisorbed molecules (atoms, ions) are neglected;
- reaction is allowed only between already chemisorbed molecules (atoms, ions) that occur at the same active site;
- the reaction products are immediately remove to the gas phase.

The physical basis of the notion of the active site is that a specific for each gas/surface system potential well is formed at certain places on the surface, called active sites. When a gas species hits the surface at an empty active site, it may lose enough kinetic energy to be trapped in one of the highly excited bound states of the well. The further relaxation to the ground (chemisorbed) state involves several steps ascertained by the energy loss mechanism specific to the system. The form of the potential well and the energy loss mechanism determines the characteristics of any relaxation. There is, however, an important general property of any relaxation: at any excited level there are four open channels. They are: the excited species can relax to a lower level, it can be scattered inelastically to a higher level, it can stay at the same level at the same site or it can migrate to a next site. The probabilities for any of these events at any level can be evaluated by a quantum-mechanical approach appropriate for the system. Further in the paper this relaxation is called ordinary relaxation. It, however, can be interrupted at any level by a chemisorbed species that reaches that active site by migration. Since no more than one species can be chemisorbed on a single active site, the *ad hoc* assumption states that the involving of another chemisorbed species in the process of relaxation changes the Hamiltonian qualitatively - from attractive it becomes repulsive. Here we encounter again the assumption about the boundedness of fluctuations: the microscopic stability of the system considered is presented by the fact that no more than one species can be chemisorbed at a single active site. The qualitative change of the Hamiltonian ensures that property and it is achieved by involving a *finite* energy (it does not exceed the adsorption energy) and *finite* matters (a single species). Therefore, the excited species can only be scattered to a higher level or migrate to another active site. The excited species can complete the relaxation only if, after migration, it reaches an empty active site. The probability for the adsorption, however, changes non-smoothly compared to the probability for the ordinary relaxation. The ordinary relaxation can be interrupted by another species at any level, since there is no special moment for that species to occur at the active site. In other words, since the relaxation and the diffusion are processes considered independent of one another, a non-perturbative interaction can happen at any level of ordinary relaxation with equal probability. Therefore, the probability for chemisorption becomes a multi-valued function, each selection of which corresponds to one of the levels at which a non-perturbative interaction is carried out. The values of the different selections are specific for any system and they can be evaluated by an appropriate quantum-mechanical approach.

According to the general statistical concept, the separation of the micro- and macro-level of description is made at the expense of assuming Markovian property. The latter states that: the probability for any transition $\lambda_{ij}$ is computed at the micro-level at the averaged over the phase space state of the system; then the probability for this transition at macro-level admits the following property:

$$P_{ij} = \lambda_{ij}\Delta t + o(\Delta t), \qquad (1)$$

where $\lambda_{ij}$ does not depend on $\Delta t$ and the probability that more than one event happens in $\Delta t$ is negligible. The property is known as ordinarity of the process [18] and it is plausible when the law of large numbers is relevant. The latter is valid far from the points of macroscopic instabilities and far from the unstable macroscopic states, which is the case considered by us. Therefore, the macroscopic equations, derived on the assumption (1), comprise probabilities, obtained at micro-level, at their r.h.s. And the probabilities appear as multi-valued functions when a diffusion-induced non-perturbative interaction is involved. The stochastic behavior arises from the permanent random choice among selections. The fluctuations that arise from that choice are bounded: each selection is a probability and so its value is confined in the range [0,1].

The number of the diffusion-induced non-perturbative interactions that a trapped species can undergo during relaxation depends only on the concentration of the chemisorbed species $n$ and the ratio $\tau_d/\tau_a$, where $\tau_d$ is the duration of a single hop and $\tau_a$ is the duration of the ordinary relaxation. In general, the dependence on both parameters is not factorized. Next the following case is considered:

$$n \ll 1,$$
$$\tau_d/\tau_a \ll 1. \qquad (2)$$

Under these conditions an expansion of the probability for series adsorption over the number of non-perturbative interactions is available. The lowest order term gives the probability for the ordinary adsorption. The term that brings about a single diffusion-induced non-perturbative interaction is the first one over $n$ and the first one over $\tau_d/\tau_a$. This term corresponds to the following case: a species is trapped in the immediate neighborhood of a chemisorbed species. Then, most probably, the trapped species undergoes a single diffusion-induced non-perturbative interaction. Thus, the probability for a single diffusion-induced non-perturbative interaction is proportional to the concentration of the chemisorbed species $n$. It is worth noting that the latter is a particular property of the open systems. In the corresponding closed system the probability for a single non-perturbative interaction cannot exceed $n^2$.

It has been found out [3] that the conditions presented by (2) ensure the adiabatic elimination of the deviations from the Poissonian spatial distribution. Therefore, far from the points of macroscopic instabilities the macroscopic evolution is described by phenomenological equations in which the probabilities for adsorption and reaction are expanded in series over the number of diffusion-induced non-perturbative interactions. To this order of expansion, these probabilities can be presented in the following way:

$$A(\vec{n}) = A_{\det}(\vec{n}) + \mu_{ai}(\vec{n}),$$
$$R(\vec{n}) = R_{\det}(\vec{n}) + \mu_{ri}(\vec{n}), \qquad (3)$$

where $\vec{n}$ is the vector of the reaction species; $A_{\det}(\vec{n})$ ($R_{\det}(\vec{n})$) are sums of the probabilities for the ordinary adsorption (ordinary reaction) and the average values of all the possible selections at the adsorption (reaction). The subscript $i$ serves to stress that only one selection, randomly chosen among all the possible, is realized at a given moment.

The definition of $\mu_{ai}(\vec{n})$ and $\mu_{ri}(\vec{n})$ determines a general property that is very important and plays a crucial role in the next considerations. It is that the values of $\mu_{ai}(\vec{n})$ and $\mu_{ri}(\vec{n})$ are bounded in the range [-1,1] and their average value is zero. The limits of the range are set by the fact that $\mu_{ai}(\vec{n})$ and $\mu_{ri}(\vec{n})$ are probabilities. The possibility for being both positive and negative is set by their definition through (3).

Thus, the macroscopic evolution of a reaction is considered by the following type of system of stochastic differential equations for the reaction mechanism:

$$\frac{d\vec{n}}{dt} = \vec{\alpha}\hat{A}_{\det}(\vec{n}) - \vec{\beta}\hat{R}_{\det}(\vec{n}) + \vec{\alpha}\hat{\mu}_{ai}(\vec{n}) - \vec{\beta}\hat{\mu}_{ri}(\vec{n}), \qquad (4)$$

where $\vec{\alpha}$ and $\vec{\beta}$ are these parts of the probabilities that explicitly depend on the external constraints and thus, for steady conditions, appear as control parameters in (4).

There is a general confinement imposed on the dimensionless form of (4). Since any term in the r.h.s. of (4) represents a probability (and thus, its value cannot exceed 1), neither the concentration $n_i$ nor any change of it $\left|\frac{dn_i}{dt}\right|$ can exceed 1. Therefore, the value of the r.h.s. of (4) is bounded in the range [-1,1] for any admissible choice of $\alpha_i$ and $\beta_i$. Since $\alpha_i$ and $\beta_i$ vary independently of $\hat{A}_{\det}(X)$, $\hat{R}_{\det}(X)$, $\hat{\mu}_{ai}(\vec{n})$ and $\hat{\mu}_{ri}(\vec{n})$, the values of the control parameters should not exceed 1 for every admissible choice. This restriction provides the unique way of determination of the dimensionless form of (4), which is ensured by choosing for this purpose the parameter of the largest possible value among all the admissible choices.

## 3. Simulation of Chaotic Dynamical Systems

Chaotic behavior of the simulated dynamical systems that possess strange attractors is a very intriguing phenomenon. On the one hand, the dynamical systems have smooth r.h.s. that fulfills Lipschitz conditions. Therefore, their solutions should be smooth curves whose evolution is determined by the initial conditions. On the other hand, a simulated phase trajectory varies irregularly confined in the so-called strange attractor. Moreover, large sensitivity towards initial conditions is observed. Such behavior is more typical for a system of stochastic equations. But can stochastisity arise at simulation? The process of simulation involves an inevitable round off at every point whose influence about stable solutions is negligible. However, when a dynamical system behaves in the above way it is characterized by at least one positive Lyaponov coefficient. Due to this, the small fluctuations that come from the round off are enhanced in course of time to the value compatible to those of the smooth parts. This effectively produces a stochastic behavior of the r.h.s. of the dynamical systems. The amplitude of the stochastic terms is bounded since each phase trajectory is confined in the region of the strange attractor. So, the mechanism of the stochastisity at the simulation is the same as at the diffusion-induced noise, namely:

(I) There are a limited number of selections at every point and only one of them is realized at that point. The selections are two at simulation: one is the round off to the higher possible value, the other one is the round off to the lower possible value;

(ii) The selection appearance has a Markovian property. At every point there is random choice between the two possible round-off values;

(iii) The amplitude of the stochastic terms is bounded.

That is why the mathematical description of the simulated dynamical systems is the same as that of the diffusion-induced noise and next it is considered by a system of the same type:

$$\frac{d\vec{X}}{dt} = \vec{\alpha}\hat{A}_{\det}(\vec{X}) - \vec{\beta}\hat{R}_{\det}(\vec{X}) + \vec{\alpha}\hat{\mu}_{ai}(\vec{X}) - \vec{\beta}\hat{\mu}_{ri}(\vec{X}), \qquad (5)$$

where $\hat{A}_{\det}(\vec{X})$, $\hat{R}_{\det}(\vec{X})$ are the smooth parts of the r.h.s. of the original dynamical system; $\vec{\alpha}$, $\vec{\beta}$ are the control parameters; $\hat{\mu}_{ai}(\vec{X})$, $\hat{\mu}_{ri}(\vec{X})$ are the stochastic terms that originate from the round off. They have the following properties: (I) their average values are zero; (ii) their amplitude is bounded; (iii) their appearance obeys Markovian property.

It is to be expected that the common properties of the stochastisity yield in certain common characteristics of diffusion established analytically the following. The stochastic terms give rise to a continuous strictly decreasing band in the power spectra, which has the same shape for both phenomena; the structure of a physical and a strange attractor is identical; the Kolmogorov entropy has positive and finite value in both cases.

## 4. Power Spectrum and Variance

The goal of the present section is to establish the shape of the power spectrum that comes from a solution of the system of stochastic equations of the type (4) and (5) in their dimensionless form:

$$\frac{d\vec{X}}{dt} = \vec{\alpha}\hat{A}_{det}(\vec{X}) - \vec{\beta}\hat{R}_{det}(\vec{X}) + \vec{\alpha}\hat{\mu}_{ai}(\vec{X}) - \vec{\beta}\hat{\mu}_{ri}(\vec{X}). \qquad (6)$$

The proof is based only on the boundedness of the solution at any instant and on the Markovian property of the stochastic terms (selections) appearance. The major idea presented in the Introduction that the thresholds of variability and the variance of fluctuations have values specific to a given system but always finite is used as well. Next it is proved that the permanent boundedness is ensured if and only if the power spectrum of any bounded irregular time series of finite length $T$ comprises a strictly decreasing band which uniformly fits the shape $1/f^{\alpha(f)}$. $\alpha(f) \in [1, p]$, $p$ is arbitrary but $p > 2$. $\alpha(f)$ is a continuous everywhere function that strictly increases starting from the value $\alpha = 1$ at $f = 1/T$ up to $\alpha = p$ as $f$ approaches infinity. In turn, this shape immediately provides a finite variance of fluctuations. The latter corresponds with the initial assumption about the boundedness of fluctuations, since any bounded sequence has finite average and finite variance. Thus, the proof is self-consistent.

The derivation of the spectral density shape follows several steps first of which is to find out that the autocorrelation function is a power one. By definition it is:

$$G(\eta) = \lim_{T \to \infty} \frac{1}{T} \int_0^T (X(t+\eta) - \overline{X})(X(t) - \overline{X}) dt, \qquad (7)$$

where $\overline{X}$ is the average value of $X(t)$. For the sake of simplicity, only one component of $\vec{X}(t)$ is considered suppressing the subscript. The generalization is trivial. The next item is the evaluation of $\overline{X}$. Assume $X(t)$ does not comprise any monotone component, then

$$\begin{aligned}\frac{d\overline{X}}{dt} &= \frac{d}{dt} \lim_{T \to \infty} \frac{1}{T} \int_0^T X(t) dt = \lim_{T \to \infty} \frac{1}{T} \int_0^T \frac{dX(t)}{dt} = \\ &= \lim_{T \to \infty} \frac{1}{T} \int_0^T (\alpha A_{det}(X) - \beta R_{det}(X)) dt + \lim_{T \to \infty} \frac{1}{T} \int_0^T (\alpha \mu_{ai}(X) - \beta \mu_{ri}(X)) dt\end{aligned} \qquad (8)$$

The second integral is zero while the first one yields $X_{det}(t)$ where $X_{det}(t)$ is the solution of:

$$\frac{d\vec{X}_{det}}{dt} = \vec{\alpha}\hat{A}_{det}(\vec{X}_{det}) - \vec{\beta}\hat{R}_{det}(\vec{X}_{det}). \qquad (9)$$

The construction of (9) and the physical foundations of its derivation imply that it serves to describe the evolution according to a deterministic law in the case of the diffusion-induced noise. That is why we called $X_{det}(t)$ deterministic component of the solution $X(t)$. In the case of the simulation (9) presents the analytical part of the solution.

As a result, when $X_{det}(t)$ is not a monotone function, $X(t)$ is an irregular bounded at each $t$ function. The invariant set designed by the admissible values of $X(t)$ is called next physical attractor $D$. Therefore, for any $t$ and any value of $X(t)$ there exists finite $u_{tMX}$ such that:

$$X(t + u_{tMX}) = M, \tag{10}$$

where $M \in D$. This property is scaling invariant with respect to averaging over a finite time period. In other words, if $\tilde{X}(\tilde{t})$ is the averaged over time period $\delta$ time series $X(t)$, there exists finite $\tilde{u}_{\tilde{t}M\tilde{X}}$ such that:

$$\tilde{X}(\tilde{t} + \tilde{u}_{\tilde{t}M\tilde{X}}) = M. \tag{11}$$

On the basis of this property the following relation holds:

$$\frac{1}{T} \int_0^{T/\delta} dt \int_t^{t+\delta} \left(X(\xi) - X_{det}\right) d\xi = \varepsilon(T) = \varepsilon(\delta)\varepsilon\left(\frac{T}{\delta}\right). \tag{12}$$

The last relation is satisfied for each $\delta$ and $T$ $(\delta, T \gg 1)$ if and only if $\varepsilon(\delta)$ is a power function. An immediate result of (12) is that the autocorrelation function $G(\eta)$ is a power function. Let us present it in the form:

$$G(\eta) = G(0)g(\eta), \tag{13}$$

where $G(0)$ is the variance of fluctuations. Our next task is to establish the explicit form of $g(\eta)$. It is to be expected that both $g(\eta)$ and its Fourier transform, power spectrum $S(f)$, are continuous strictly decreasing functions since $(X(t) - X_{det})$ is an irregular function which does not comprise any special period(s). Our assumption is that the stochastic part $(X(t) - X_{det})$ cannot give rise to any special period(s) because there are not physical foundations for distinguishing any of the periods that originate from $(X(t) - X_{det})$. Next it is assumed that $S(f)$ is a power function of the type:

$$S(f) \propto \frac{1}{f^{\alpha(f)}}, \tag{14}$$

where the explicit shape of $\alpha(f)$ is to be established. This is done on the basis of the permanent boundedness of $X(t)$ by using the Wiener-Khinchin theorem. The latter states:

$$S(f) = \lim_{T \to \infty} \frac{1}{T} \left\langle |y_T(f)|^2 \right\rangle, \qquad (15)$$

where $y_T(f)$ is the Fourier transform of $(X(t) - X_{det})$ over a period of length $T$. For any finite $T$ the Fourier spectrum is a discrete one. $\frac{1}{T}|y_T(f)|^2$ uniformly fits the shape of $S(f)$ as $T$ approaches infinity. So, $X_T(t)$ is majored by:

$$I_T(t) = \sqrt{T} \int_{1/T}^{\infty} \frac{\cos ft}{f^{\alpha(f)/2}} df + \sqrt{T} \int_{1/T}^{\infty} \frac{\sin ft}{f^{\alpha(f)/2}} df, \qquad (16)$$

where the Fourier coefficients are constructed on the basis of (14) and (15). In order for $I_T(t)$ to serve as an estimate of $X_T(t)$ it should be finite for any $t$. For this purpose it is enough to find out at what values of $\alpha(f)$:

$$I_T(0) = \max|I_T(t)| = \sqrt{T} \int_{1/T}^{\infty} \frac{1}{f^{\alpha(f)/2}} df \qquad (17)$$

is finite and its value does not depend on $T$.

The integration of a power function with a non-constant power involves the following steps: (I) the integration interval $[a,b]$ is divided into subintervals of equal length $\varepsilon$ and the value of $\alpha(f)$ is set constant equal to its value at the left point of each subinterval. (ii) By the use of standard calculus it is found out that when $\alpha(f)$ is a continuous everywhere function with a finite everywhere first derivative:

$$\int_a^b \frac{df}{f^{\alpha(f)}} = \frac{b^{1 \pm \alpha(b)}}{1 \pm \alpha(b)} - \frac{a^{1 \pm \alpha(a)}}{1 \pm \alpha(a)}. \qquad (18)$$

Eq. (18) does not change if $\alpha(f)$ intersects 1 at a certain point. Thus, if $\alpha(f)$ is an everywhere continuous function, the power spectrum $S(f)$ is also a continuous everywhere function and thus it does not comprise any "special" period(s). This satisfies our assumption that the stochastic part of $X(t)$ cannot be the onset of any special period(s). That is why, it is supposed next that $\alpha(f)$ is a continuous everywhere function with a finite first derivative.

Now we can evaluate the range of values of $\alpha(f)$, which should be such that $I_T(0)$ from (17) is finite and therefore it does not depend on $T$. The formal expression of this property is through:

$$I_T(0) = \sqrt{T} \int_{1/T}^{\infty} \frac{df}{f^{\alpha(f)/2}} \propto T^0. \qquad (19)$$

Eq. (19) holds when $\alpha(f)$ is a continuous strictly increasing function between the following limit values:

$$\alpha\left(\frac{1}{T}\right) = 1,$$
$$\alpha(\infty) = p$$
(20)

where $p$ is arbitrary but $p > 2$.

An important immediate result of (20) is that the variance of fluctuations, calculated on the ground of the spectral density, is finite. By definition, it is:

$$Var = G(0) \propto \int_0^\infty \frac{df}{f^{\alpha(f)}}.$$
(21)

Since the integral in (21) is improper, its value is determined trough:

$$Var = G(0) = \lim_{T \to \infty} Var_T = \int_{1/T}^\infty \frac{df}{f^{\alpha(f)}},$$
(22)

where $Var_T$ is the variance of fluctuations in a time interval of finite length $T$. The meaning of (22) is that $Var_T$ uniformly approaches $Var$ as $T \to \infty$. It is obviously that when $\alpha(f)$ is a continuous everywhere strictly increasing function in the range given by (20), then:

$$Var_T = \int_{1/T}^\infty \frac{df}{f^{\alpha(f)}} \propto T^0,$$
(23)

and hence $Var$ is finite.

Thus, our proof is self-consistent: the shape of the power spectrum is a result of the permanent boundedness of the variability of $X(t)$. As a result, the variance of fluctuations, calculated on the ground of this shape, is finite. And the latter is coherent with the idea about the boundedness of fluctuations since any bounded sequence has finite mean and finite variance.

The greatest advantage of the obtained power spectrum shape ($1/f^{\alpha(f)}$) is that it ensures a finite variance of the fluctuations. The major difference between the present shape and the $1/f^\alpha$ ($\alpha$ constant) is that the latter yields infinite variance no matter what the value of $\alpha$ is. Therefore it cannot ensure a stable long-term evolution of either physical system. On the contrary, the established above shape provides a long-term stable evolution of the physical systems. Another important result is that the shape $1/f^{\alpha(f)}$ is the same for any particular length of the time series. Practically, at low frequencies this shape is $1/f$-like with high accuracy because the deviations of the non-constant exponent $\alpha(f)$ from 1 are very small and become smaller and smaller as the frequency tends to $1/T$. Therefore, the $1/f$ behavior at low frequencies appears as a kind of alias effect.

Another obvious result is that the power spectrum at diffusion-induced noise can comprise both a discrete and a continuous band. The discrete band originates from the deterministic part (9) when its solution is a stable periodic function. On the contrary, the power spectra of the simulated chaotic dynamical systems cannot comprise a discrete band because there is no periodic function among their admissible analytical solutions. As a result, the superimposing (or its absence) of a discrete and the continuous band appears as a practical discrimination criterion between the diffusion-induced noise and simulated chaotic dynamical systems. More generally, the set of admissible solutions of the deterministic part (9) is the only source for different behavior. On the contrary, the stochastic part of the solutions exhibits the same behavior; namely it gives rise to a persistent continuous band of the shape $1/f^{\alpha(f)}$ in the power spectra at both phenomena. An important feature of this band is that it is insensitive to the properties of the solution of the deterministic part.

**5. Correlation Dimension and Kolmogorov Entropy**

Since the mathematical description of the stochasticity is identical for the diffusion-induced noise and the simulated chaotic dynamical systems, it is to be expected that all the properties whose onset is the stochasticity are the same. It has been already confirmed for the shape of the continuous band in the power spectra. The goal of the present section is to deduce that the two other properties, the non-integer correlation dimension and finite Kolmogorov entropy, are also shared by both phenomena. Moreover, it is established that a physical (strange) attractor is non-homogeneous and the explicit law of population is derived.

**5.1 *Correlation Dimension***

The solution $\vec{X}(t)$ of (6) constitutes a $d$ - dimensional vector field confined in the physical (strange) attractor whose embedding dimension is also $d$. The law of the phase space density determines the correlation dimension. If the phase space is randomly populated, the density is proportional to $l^{-d}$, where $l$ is the size of a small cell anywhere in the phase space. Then the correlation dimension is $d$. Next it is shown that the physical (strange) attractor is non-homogeneous and that the correlation dimension at any point deviates from $d$. The density at a given point of the attractor is proportional to the averaged over the initial values and initial moments probability for reaching that point. According to the construction of the r.h.s. of eq. 6, reaching the higher values of any $X_i(t)$ is less probable since it takes more steps (respectively time). Therefore, it is to be expected that the density of the physical (strange) attractor strictly decrease when reaching the boundary. The explicit law of this dependence is presented next. It is assumed that the value of $X_i$ serves as a measure of the distance to the boundary. Since reaching the higher value of $X_i$ takes more time, we put forth the assumption that there is a diffeomorfism between the averaged over the initial values and the initial moments probability for reaching a given value of $X_i$ and the averaged over the initial values and initial moments time for reaching that value. The averaged time for reaching a given $X_i$ is expressed by the autocorrelation function $G_T(t)$ of $I_T(t)$ that is:

$$P_t = G_T(t) \propto \lim_{T \to \infty} \left(1 - \left(\frac{t}{T}\right)^{v\left(\frac{t}{T}\right)}\right), \tag{24}$$

where

$$\begin{aligned} v(\tau) &\to p-1 \text{ as } \tau \to 0, \\ v(\tau) &\to 0 \text{ as } \tau \to 1. \end{aligned} \tag{25}$$

The assumption about the diffeomorfism between $P_t$ presented by (24) and the averaged probability for reaching given $X_i$ yields:

$$P_x(X_i) \propto \left(1 - |X_i|^{\nu(|X_i|)}\right), \qquad (26)$$

where $\nu(\tau)$ satisfies (25)

$$\begin{aligned} \nu(|X_i|) &\to p-1 \quad \text{as} \quad |X_i| \to 0, \\ \nu(|X_i|) &\to 0 \quad \text{as} \quad |X_i| \to 1. \end{aligned} \qquad (27)$$

It is worth noting that all $X_i$ are renormalised so that each of them is confined in the region [-1,1]. The same range of variability of $\frac{t}{T}$ and each $X_i$ ensures the formal coincidence of the values of $\nu\left(\frac{t}{T}\right)$ and $\nu(|X_i|)$.

Therefore, the density of the point $X_i$ in the $d$-dimensional phase space is:

$$n \propto \prod_{i=1}^{d} P_x(X_i) = \prod_{i=1}^{d} \left(1 - |X_i|^{\nu(|X_i|)}\right). \qquad (28)$$

It is obvious that effectively $n \propto l^{-\mu(\vec{X})}$ with $d \le \mu(\vec{X}) < \infty$ for each value of $X_i$ apart from the single point $X_i = 0$ where $\mu(0) = d$. Hence, the physical (strange) attractor is non-homogeneous.

An important immediate result of (24) and (26) is that the probability for reaching any point in the physical attractor for finite time starting anywhere in the attractor is finite. This constitutes ergodicity of the process that brings about the diffusion-induced noise introduced in &2. Ergodicity along with the Markovian property of the selection appearance (stochastic terms in (4)) ensures the frame of the general concept of the statistical mechanics, which states that the fluctuations on macro-level should be presented by an ergodic Markovian process. The major difference between this process and the Ornstein-Uhlenbeck one is that the latter requires an asymptotically stable single averaged value. The ergodicity and the Markovian property of the diffusion-induced noise are insensitive to the features of the deterministic solution of (6), determined by (9), when it is a stable one.

**5.2 *Kolmogorov Entropy***

By definition Kolmogorov entropy is the average rate of losing information about the evolution of phase trajectories. The notion of information entropy is related to the name of Shannon [19]. According to him, the average rate of losing information is given by:

$$K = -P(\vec{X}(\eta)) \ln P(\vec{X}(\eta)), \qquad (29)$$

where $P(\vec{X}(\eta))$ is the averaged over the phase trajectory probability that a given trajectory deviates from the point $\vec{X}$ in a time interval $\eta$. Next the explicit form of $P(\vec{X}(\eta))$ is established. According to its definition, the value of the autocorrelation function $G(\eta)$ gives the average probability that two points in the time series of $X_i(t)$ separated by time interval $\eta$, have the same value. On the other hand, the average probability that this value is $X_i$ is given by $P_x(X_i)$ from (26). In other words, the average probability that a given trajectory intersects itself at the point $X_i$ in time interval of length $\eta$ is:

$$P_{x\eta}(X_i(\eta),\eta) \propto \frac{1-|X_i|^{\nu(|X_i|)}}{\widetilde{\eta}^{\rho(\eta)}}, \qquad (30)$$

where: $1/\eta^{\rho(\eta)}$ is $g(\eta)$ from (13). Its explicit shape can be evaluated as an inverse Fourier transform of the power spectrum $S(f)$. $\widetilde{\eta}$ has the following meaning: since the stochastic terms in (6) have finite values, it takes finite time for realization of a self-intersection. It is assumed that there exists $\eta_0$ such that the probability for self-intersection at $\eta < \eta_0$ is zero. Then, $\widetilde{\eta}$ is defined through $\widetilde{\eta} = \frac{\eta}{\eta_0}$. Thus, $\widetilde{\eta}$ is always greater than 1 which ensures that the denominator is always greater than 1. In other words, $P_{x\eta}(X_i(\eta),\eta)$ is the probability for predicting the position of the phase trajectory in a given time interval. Then, $(1 - P_{x\eta}(X_i(\eta),\eta))$ is the probability for any deviation from the predictability. Then:

$$K = -\left(1 - \prod_{i=1}^{d}\prod_{\xi=0}^{\infty}\prod_{\zeta=1}^{\infty}\frac{1-|X_{i\xi}|^{\nu(|X_{i\xi}|)}}{\widetilde{\eta}_\zeta^{\rho(\eta_\zeta)}}\right) \ln\left(1 - \prod_{i=1}^{d}\prod_{\xi=0}^{\infty}\prod_{\zeta=1}^{\infty}\frac{1-|X_{i\xi}|^{\nu(|X_{i\xi}|)}}{\widetilde{\eta}_\zeta^{\rho(\eta_\zeta)}}\right), \qquad (31)$$

where the subscript $\xi$ takes all the values in the range $[0,\infty)$. This means that the physical attractor is divided into cells of infinitesimally small size. The subscript $\zeta$ starts from 1 in order to stress that $\widetilde{\eta}$ starts from 1. The partition is made in order to provide averaging over different phase trajectories. It is obvious that for any finite length time series and for any partition into finite size cells $K$ from (31) is positive and finite. This is a surprising result since the physical attractor originates from a random process and, so far, it is accepted that $K$ should be positive but *infinite* for any random process [20,21].

**Conclusions**
      The general hypothesis of the present paper is about the influence of the boundedness of the fluctuations. The physical notion of the boundedness of fluctuations asserts that a long-term stable evolution of a given system is provided when the fluctuations are driven by involving locally only *finite* energy and/or matter so that the system permanently stays within the thresholds of the stability. However, the idea about the boundedness of fluctuations is more general since it can arise due to variety of reasons. We consider two subjects with bounded fluctuations where the boundedness is of different origin. These are (I) a new physical mechanism that brings about large-scaled fluctuations (noise) at stable states. The

origin of boundedness is that involving locally finite energy and matter so that the system stays microscopically stable drives any fluctuation. (ii) The fluctuations induced by simulation at chaotic dynamical systems. The boundedness is a result of the structure of some differential equations such that their solutions are confined in a strange attracttor. It is established that the stochasticity that arises in both cases has identical properties. So, it is to be expected that it be manifested in certain common features of both types of systems. Indeed, it is found out that the power spectrum of any bounded irregular time series of length $T$ comprises a strictly decreasing band that uniformly fits the shape $1/f^{\alpha(f)}$ where $\alpha\left(\frac{1}{T}\right)=1$ and $\alpha(f)$ tends to $p>2$ as $T$ approaches infinity. Practically, this shape is $1/f$ like with high accuracy at low frequencies because the deviations of the non-constant $\alpha(f)$ from 1 are small and become even smaller as the frequency approaches $1/T$. The greatest advantage of this shape is that it provides a finite variance of fluctuations. Thus, reconciliation between the $1/f^\alpha$-type tails in the power spectra and the long-term stable evolution is achieved. It is worth noting that the established shape is a result of the boundedness of the fluctuations. It is worth noting also that the major difference between the shape $1/f^{\alpha(f)}$ and the $1/f^\alpha$ ($\alpha$ constant) one is that the latter yields infinite variance of fluctuations no matter what the value of $\alpha$ is. Therefore, it cannot provide stable long-term evolution of any physical system.

Next it is found out that the structure of the physical and strange attractors is identical, the correlation dimension is non-integer, the Kolmogorov entropy is positive and finite. As a result, these properties cannot serve as discrimination criteria between both phenomena.

We establish that the only onset of discrimination lies in the properties of the deterministic part of (6) presented by (9). There are no limit cycles among the admissible solutions of (9) at the chaotic dynamical systems whereas they are not banned at the diffusion-induced noise. As a result, the superimposing (or its absence) of a discrete and the continuous band appears as a practical discrimination criterion between the diffusion-induced noise and simulated chaotic dynamical systems.

Another important result is that the diffusion-induced noise is an ergodic Markovian process whose major peculiarity is that this property is insensitive to the mathematical characteristics of (9) if only the latter is a stable one.

**References**

1. I. Birol, A. Hacinlyan and Z. Perdahci, "*Simulating a dynamical system in the presence of $1/f$ noise*", *Int. J. Engng. Sci.*, **36** (1998) 265–271.
2. M. K. Koleva, A. E. Elyias and L. A. Petrov, "*Fractal power spectrum at catalytic oxidation of HCOOH over supported Pd catalyst*" in Metal-Ligand Interactions in Chemistry, Physics and Biology, NATO ASI Series C, Vol. 546, eds. N. Russo and D. R. Salahub, Kluwer Academic Publishers, Dordrecht, 2000, p. 353–369.
3. M. K. Koleva, "*Non-perturbative interactions: a source of a new type noise in open catalytic systems*", *Bulg. Chem. Ind.*, **69** (1998) 119–128.
4. W. Horsthemke and R. Lefever, *Noise-Induced Transitions*, Springer-Verlag, Heidelberg, (1984).
5. A. I. Olemskoi, *Teoriya stokhasticheskih system s singulyarnyim multiplicativnyim shumom*, *Usp. Fiz. Nauk*, **168** (1998) 287–321.
6. Yu. L. Klimontovich, *Nelineyinoe brounovskoe dvizenie*, *Usp. Fiz. Nauk*, **164** (1994) 811–843.

7. G. Nicolis, "*Some aspects of fluctuation theory in non-equilibrium systems*", in Stochastic NonLinear Systems in Physics, Chemistry and Biology*,* Eds. L. Arnold and R. Lefever Springer-Verlag, Berlin-Heidelberg, 1981, p. 44–52.
8. N. G. van Kampen, *Stochastic Processes in Physics and Chemistry*, North Holland, Amsterdam (1981).
9. G. Nicolis and I. Prigogine, *Self-Organization in Non-Equilibrium Systems*, Wiley, New York (1977).
10. (a) C. Knessl, M. Mangel, B. J. Matkovsky, Z. Schuss and C. Tier, "*Solutions of Kramers-Moyal equations for problems in chemical physics*", *J. Chem. Phys*., **81** (1994) 1285–1293.
11. R. Graham and T. Tel, Phys. Rev. A, "*Weak-noise limit of Fokker-Planck models and non-differentiable potentials for dissipative dynamical systems*", *Phys. Rev. A*, **31** (1985) 1109–1122.
12. V. A. Chinarov, M. I. Dykman and V. N. Smelyanskiy, "*Dissipative corrections to escape probabilities of thermal nonequilibrium systems*", *Phys. Rev. E*, **47** (1993) 2448–2461.
13. A.D. Wentzell and M.I. Freidlin, "*Fluctuations in Dynamical Sysytems under the Action of Small Random Perturbations*", Nauka, Moscow, 1979 (in Russian); English translation "*Random Perturbations in Dynamical Systems*", Springer, New York, (1984).
14. M. I. Dykman, E. Mori, J. Ross and P. M. Hunt, *Large fluctuations and optimal paths in chemical kinetics*, *J. Chem. Phys*., **100** (1994) 5735–5750.
15. C.W. Gardiner, *Handbook of Stochastic Methods*, 2-nd Ed., Springer, New York, (1985) (c) Yu. L. Klimontovich, "Statisticheskaya Fisika", Mir, Moscow, 1982 (in Russian).
16. L. S. Polak and A. S. Mikhailov, "*Self-Organization in Non-Equilibrium Physico-Chemical Systems*", Nauka, Moscow (1983) (in Russian).
17. Yu. L. Klimontovich, "*Statistical Physics*", Mir, Moscow (1982) (in Russian).
18. B. V. Gnedenko, "*A Course of Probability Theory*", Nauka, Moscow (1988) (in Russian).
19. C. E. Shannon and W. Weaver, "*The Mathematical Theory of Information*", University of Illinois Press, Urbana (1949).
20. H. G. Schuster, "*Deterministic Chaos. An Introduction*", Physik-Verlag, Weinheim, (1984).
21. D. Ruelle, "*Deterministic Chaos: The Science and Fiction*", *Proc. Roy. Soc. London*, **A427** (1990) 241–248.